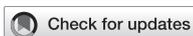

# Mechanisms of SiO oxidation: Implications for dust formation

Stefan Andersson[1,2]*, David Gobrecht[1] and Rosendo Valero[3,4]

[1]Department of Chemistry and Molecular Biology, University of Gothenburg, Gothenburg, Sweden, [2]SINTEF, Trondheim, Norway, [3]Department of Chemistry, University of Coimbra, Coimbra, Portugal, [4]Institute for Frontier Materials, Zhejiang Huayou Cobalt Co. Ltd, Tongxiang, Zhejiang, China

Reactions of SiO molecules have been postulated to initiate efficient formation of silicate dust particles in outflows around dying (AGB) stars. Both OH radicals and $H_2O$ molecules can be present in these environments and their reactions with SiO and the smallest SiO cluster, $Si_2O_2$, affect the efficiency of eventual dust formation. Rate coefficients of gas-phase oxidation and clustering reactions of SiO, $Si_2O_2$ and $Si_2O_3$ have been calculated using master equation calculations based on density functional theory calculations. The calculations show that the reactions involving OH are fast. Reactions involving $H_2O$ are not efficient routes to oxidation but may under the right conditions lead to hydroxylated species. The reaction of $Si_2O_2$ with $H_2O$, which has been suggested as efficient producing $Si_2O_3$, is therefore not as efficient as previously thought. If $H_2O$ molecules dissociate to form OH radicals, oxidation of SiO and dust formation could be accelerated. Kinetics simulations of oxygen-rich circumstellar environments using our proposed reaction scheme suggest that under typical conditions only small amounts of $SiO_2$ and $Si_2O_2$ are formed and that most of the silicon remains as molecular SiO.

## 1 Introduction

The SiO molecule has been observed in the interstellar medium and in stellar outflows and is believed to be important for the formation of interstellar dust, which to a large extent consists of silicates (Hartquist et al., 1980; Clegg et al., 1983; Herbst et al., 1989; Langer and Glassgold, 1990; Sternberg and Dalgarno, 1995; Schilke et al., 1997; Gail and Sedlmayr, 1999; Smith et al., 2004; Gusdorf et al., 2008; Reber et al., 2008; Goumans and Bromley, 2012; Chakraborty et al., 2013; Plane, 2013; Krasnokutski et al., 2014; Bromley et al., 2016; Gobrecht et al., 2016). SiO is also found in terrestrial environments such as the upper atmosphere (from meteoric ablation; see Plane et al., 2016), in combustion of silicon compounds (Jachimowski and McLain, 1983; Britten et al., 1990; Tokuhashi et al., 1990; Chagger et al., 1996; Lindackers et al., 1997; Wooldridge, 1998; Moore et al., 2006) and in industrial silicon production processes (Johansen et al., 1998; Schei et al., 1998; Ravary and Johansen, 1999; Grådahl et al., 2007; Ravary et al., 2007; Kamfjord et al., 2012; Næss et al., 2014). SiO can react with oxygen-bearing species, to form $SiO_2$ (Gómez Martín et al., 2009; Chakraborty et al., 2013). The reactions

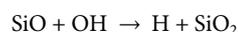

$$SiO + OH \rightarrow H + SiO_2$$

and

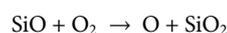

$$SiO + O_2 \rightarrow O + SiO_2$$

have been suggested to be important for gas phase $SiO_2$ formation in various environments. In contrast to its carbon analogue, $CO_2$, the thermodynamically most stable state of $SiO_2$ is a condensed phase, silica, at ambient conditions. Molecular $SiO_2$ will readily form $SiO_2$ clusters and eventually condense to silica particles at sufficiently high concentrations (in sufficiently dense media; see Lindackers et al., 1997; Wooldridge, 1998; Næss et al., 2014). SiO could also nucleate directly to form clusters, and through a series of reactions with different species form silica or silicate dust (Reber et al., 2008; Goumans and Bromley, 2012; Plane, 2013; Krasnokutski et al., 2014; Bromley et al., 2016). There is a lack of experimental data on the elementary reactions involved in SiO oxidation, with a few exceptions (Gómez Martín et al., 2009), and computational chemistry techniques are essential in providing mechanistic insight and estimates of the rates of reaction (Zachariah and Tsang, 1995; Becerra et al., 2005; Reber et al., 2008; Gómez Martín et al., 2009; Goumans and Bromley, 2012; Plane, 2013; Hao et al., 2014; Bromley et al., 2016; McCarthy and Gauss, 2016; Yang et al., 2018).

Experimentally, $SiO_2$ formation has commonly been inferred from decay of SiO reactants in reactions with, e.g., $H_2O$, OH, $O_2$, and $O_3$ (Gómez Martín et al., 2009). Molecular $SiO_2$ products have proven harder to observe but have been detected directly in a few studies by time-of-flight mass spectrometry, for instance Yang et al. (2018) by electron-impact ionization of products of the SiH + $O_2$ reaction and by Kostko et al. (2009) who generated gas phase $SiO_2$ by laser ablation of silicon in a $CO_2$ molecular beam followed by photoionization.

In space, SiO has been observed most abundantly in outflows in star forming regions where temperatures range from 10 K to 2000 K (Hartquist et al., 1980; Clegg et al., 1983; Schilke et al., 1997; Gusdorf et al., 2008). The SiO abundance increases rapidly with increasing temperatures. SiO is assumed to be formed mainly through the shock-induced evaporation of nano-to micrometer sized particles, "interstellar dust grains", which to a large extent consist of silicates. SiO can be formed either directly from the grains or through sputtering of Si atoms that subsequently react with $O_2$ and/or OH. The main gas-phase mechanism for removal of SiO is assumed to be the reaction with OH. The rate of this reaction is highly uncertain, especially at low temperatures.

A large portion of interstellar dust is assumed to be formed in the outflow from Asymptotic Giant Branch (AGB) stars, i.e., dying stars in a red giant phase (Gail and Sedlmayr, 1999; Gail et al 2013). Since the first observations of circumstellar silicate emission bands (Woolf and Ney, 1969; Hackwell et al., 1970) the formation of silicate dust grains represents a challenging problem for both theory and experiments. The theoretical study of Gail and Sedlmayr (1998) based on classical nucleation theory and thermodynamic equilibrium provoked a vibrant discussion in the community (Ali et al., 1998). This discussion points out the importance of non-equilibrium processes, the drawbacks of steady-state rates used in classical homogeneous nucleation theory, and the non-crystalline character of small cluster structures. This is in line with the chemical-kinetic bottom-up approach of the present study.

In oxygen-rich AGB stars the formed dust is mainly in the form of silicates, containing either magnesium or iron or both, and show an olivine-type [general formula $(Mg,Fe)SiO_4$] or pyroxene-type [general formula $(Mg,Fe)SiO_3$] stoichiometry. In AGB outflows, most circumstellar silicates are found to be magnesium-rich and iron-poor (Woitke, 2006; Höfner, 2008). The energies and structures of the most favourable Mg-rich olivine and pyroxene clusters were investigated in a recent study (Macià Escatllar et al., 2019). However, the formation of the corresponding monomer represents a major challenge in modeling the nucleation of astrophysical silicate dust.

Several stepwise mechanisms have been suggested as responsible for dust formation in these environments, where typical physical conditions are pressures of 0.001–0.1 Pa and temperatures of 1,000–1200 K (Goumans and Bromley, 2012; Plane, 2013; Bromley et al., 2016; Gobrecht et al., 2016). Based on density functional theory (DFT) calculations, Goumans and Bromley (2012) suggested a seven-step mechanism starting with the two reactions:

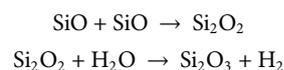

$$Si_2O_2 + H_2O \rightarrow Si_2O_3 + H_2$$

The latter reaction is exothermic and expected to be efficient even though the kinetics of the reaction are unknown. In this mechanism, $Si_2O_3$ is a key species, which, once formed, can lead to efficient growth of larger particles in reaction with other species, such as $H_2O$ and Mg atoms. Reaction with $H_2O$ would lead to formation of a hydroxylated species:

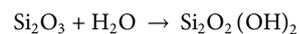

Cluster beam experiments were performed to investigate the agglomeration of SiO molecules (Reber et al., 2006). They revealed atomic segregation of silicon for $(SiO)_n$ clusters with sizes larger than n = 7. More recently, this size-dependent trend of silicon segregation in $(SiO)_n$ clusters was confirmed by Bromley et al. (2016). They investigated the kinetics of the nucleation of SiO into large $(SiO)_n$ clusters using DFT (B3LYP) to calculate structures and energetics. SiO nucleation had previously been suggested as a viable reaction route for initiating dust formation (Reber et al., 2006; Reber et al., 2008). However, this was found to not be an efficient process under the low pressures in the circumstellar environments studied, and only the smallest clusters, $Si_2O_2$ and $Si_3O_3$, were found to form in any significant amount during a reasonable time scale.

Small silicon oxide species like $SiO_2$, $Si_2O_3$, and $Si_2O_4$ show strong absorption features between 7.0 μm and 8.0 μm (Gardner et al., 2020). These species, if present, could be detected with the observing facilities of the Atacama Large Millimeter/submillimeter Array (ALMA) and the James Webb Space Telescope (JWST).

Stellar dust formation is fundamental for the galactic chemical evolution as it is responsible for the vigorous mass loss of late-type stars enriching the interstellar medium. The formed silicate grains are later important as providing a catalytic surface for the initial formation of most molecules observed in space, including complex organic molecules, and as providing the solid material that makes up the dusty disks around newly-formed stars eventually creating planetary systems.

Based on the previous work, we decided to study the kinetics of the oxidations and clustering reactions leading to growth of silicon oxide molecules with up to three Si atoms. The reported work does to a large extent build upon parallel studies of the energetics of the reactions of SiO and $Si_2O_2$ with OH and $H_2O$ (Andersson, to be published).

This paper is structured as follows. First, we describe the computational methods used. Then we present results on master equation calculations of the rate coefficients of relevant reactions. This is followed by a description of the kinetic model. Subsequently, kinetic simulation results are presented on models of circumstellar regions of oxygen-rich AGB stars. Finally, we present some concluding remarks and a summary of the most important findings of this study.

## 2 Computational details

For optimizing molecular structures and calculating vibrational frequencies of species involved in the reactions under study we have used the M06 density functional (Zhao and Truhlar, 2008a; Zhao and Truhlar, 2008b) with the maug-cc-pV (T+d)Z basis sets (Papajak and Truhlar, 2010). This approach was benchmarked against highly accurate CCSD(T) calculations for Si-O-H molecules and was shown to offer accurate energetics at a reasonable computational cost (Andersson, to be published). The mean deviations of M06 energies for stationary points on the potential energy surface for the SiO + OH and SiO + $H_2O$ reactions were 1.8 kJ/mol and 6.2 kJ/mol, respectively. In an earlier study by Ma et al. (2019), heats of formation calculated by M06/maug-cc-pV(T+d)Z were within the experimental error bars for SiO, $SiO_2$, $Si_2O_2$, and $SiO(OH)_2$ and deviated by only 10 kJ/mol from the experimental value for $Si(OH)_4$. All DFT calculations were performed using the NWChem program package (Valiev et al., 2010).

For the calculation of rate coefficients, Rice–Ramsperger–Kassel–Markus (RRKM) calculations were employed using the Master Equation Solver for Multi-Energy well Reactions (MESMER) program (Glowacki et al., 2012). These calculations are run in a similar fashion as described in Gobrecht et al. (2022). The energetics, rotational constants and vibrational frequencies for molecules and intermediates were taken partly from a separate study (Andersson, to be published) but calculations on the reactions of $Si_2O_3$ with OH and $H_2O$ as well as clustering reactions of silicon oxide molecules were performed in the current study. Intermediate structures on the potential energy surfaces of the $Si_2O_3$ + $H_2O$ and $Si_2O_3$ + OH reactions are given in Supplementary Tables S1, S2, respectively. Since the potential energy surfaces (PESs) of some reactions are highly complex, especially the reactions with $H_2O$, the reactions were simplified, ignoring high-energy pathways (likely to have insignificant rates) as well as lumping together intermediates with similar structures and energetics into single species (see Section 3.1). Intermediates are assumed to either dissociate back to reactants or to proceed to form products or be collisionally stabilized by background gas, in this case taken to be $H_2$. The internal energies of each intermediate were divided into a contiguous set of grains with a typical width of 110 $cm^{-1}$ containing a bundle of rovibrational states. Each grain was assigned a set of microcanonical rate coefficients for dissociation into reactants or products. The rate coefficients were determined using an inverse Laplace transform procedure to connect them to capture rate coefficients calculated using long-range transition state theory (Georgievskii and Klippenstein, 2005), in the cases where there is no potential energy barrier to reaction. The collisional transfer probability was estimated using the exponential down model, with the average energy for downward transitions designated $<\Delta E_{down}>$ and the upward transition probability determined by detailed balance. $<\Delta E_{down}>$ was assumed to be temperature independent and assigned a value of 200 $cm^{-1}$, appropriate for $H_2$ (Gilbert and Smith, 1990). The master equation, which describes the time evolution of grain populations, was expressed in matrix form and solved to yield rate coefficients of bimolecular and recombination reactions at given temperatures and pressures.

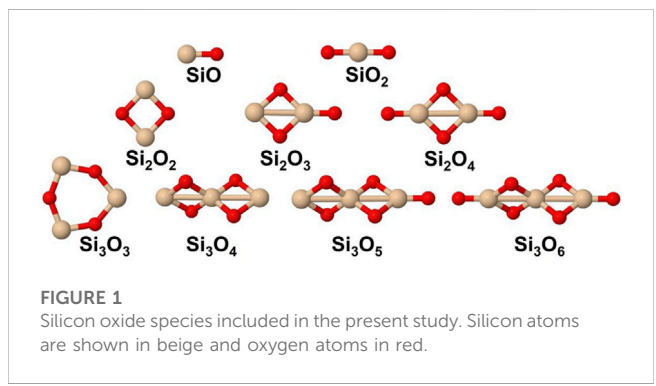

FIGURE 1
Silicon oxide species included in the present study. Silicon atoms are shown in beige and oxygen atoms in red.

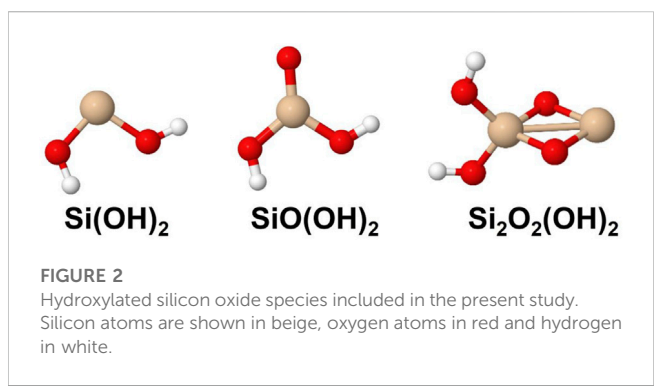

FIGURE 2
Hydroxylated silicon oxide species included in the present study. Silicon atoms are shown in beige, oxygen atoms in red and hydrogen in white.

## 3 Results and discussion

### 3.1 Rate calculations

The set of small silicon oxide molecules (or clusters) that were considered in this study are shown in Figure 1 and hydroxylated species are shown in Figure 2. Two main types of reactions that are expected to be important in the initial stages of the growth of silica and silicate particles in AGB outflows starting from SiO, i.e., oxidation reactions where $H_2O$ and OH are the oxygen-bearing species as well as clustering reactions of the silicon oxide molecules. The included oxidation reactions are the following:

$$SiO + H_2O \rightarrow SiO_2 + H_2 \quad (1a)$$
$$SiO + H_2O + M \rightarrow Si(OH)_2 + M \quad (1b)$$
$$SiO_2 + H_2O + M \rightarrow SiO(OH)_2 + M \quad (2)$$
$$Si_2O_2 + H_2O \rightarrow Si_2O_3 + H_2 \quad (3a)$$

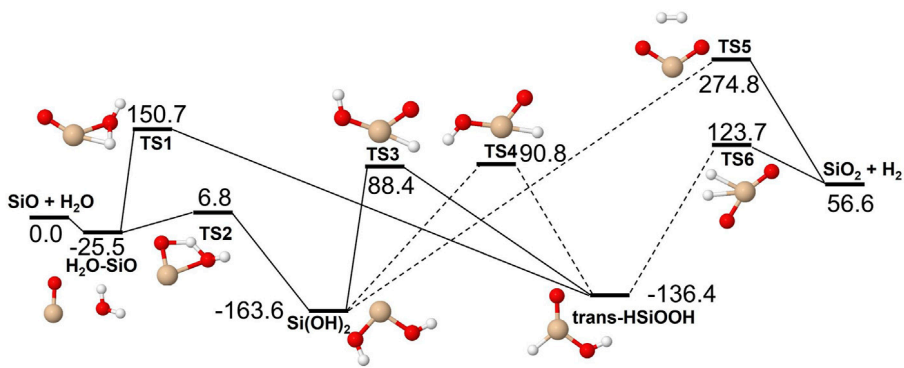

FIGURE 3
Potential energy surface based on M06 calculations for the SiO + H$_2$O reaction as implemented in the RRKM calculations. The energies are given in kJ/mol and include vibrational zero-point energy corrections. Dashed lines indicate that the reaction step has been simplified by combining several similar intermediate isomers.

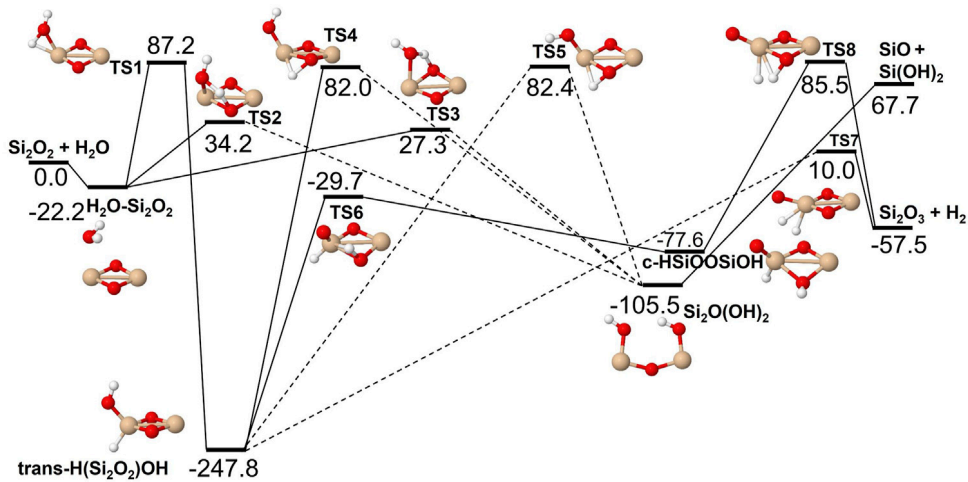

FIGURE 4
Potential energy surface based on M06 calculations for the Si$_2$O$_2$ + H$_2$O reaction as implemented in the RRKM calculations. The energies are given in kJ/mol and include zero-point energy corrections. Dashed lines indicate that the reaction step has been simplified by combining several similar intermediate isomers.

$$Si_2O_2 + H_2O \rightarrow Si(OH)_2 + SiO \quad (3b)$$
$$Si_2O_3 + H_2O \rightarrow Si_2O_4 + H_2 \quad (4a)$$
$$Si_2O_3 + H_2O \rightarrow SiO(OH)_2 + SiO \quad (4b)$$
$$Si_2O_3 + H_2O + M \rightarrow Si_2O_2(OH)_2 + M \quad (4c)$$
$$SiO + OH \rightarrow SiO_2 + H \quad (5)$$
$$Si_2O_2 + OH \rightarrow Si_2O_3 + H \quad (6)$$
$$Si_2O_3 + OH \rightarrow Si_2O_4 + H \quad (7)$$

The clustering reactions are the following:

$$SiO + SiO + M \rightarrow Si_2O_2 + M \quad (8)$$
$$SiO + SiO_2 + M \rightarrow Si_2O_3 + M \quad (9)$$
$$SiO_2 + SiO_2 + M \rightarrow Si_2O_4 + M \quad (10)$$
$$SiO + Si_2O_2 + M \rightarrow Si_3O_3 + M \quad (11)$$
$$SiO + Si_2O_3 + M \rightarrow Si_3O_4 + M \quad (12)$$
$$SiO_2 + Si_2O_2 + M \rightarrow Si_3O_4 + M \quad (13)$$
$$SiO + Si_2O_4 + M \rightarrow Si_3O_5 + M \quad (14)$$
$$SiO_2 + Si_2O_3 + M \rightarrow Si_3O_5 + M \quad (15)$$
$$SiO_2 + Si_2O_4 + M \rightarrow Si_3O_6 + M \quad (16)$$

In what follows the results of rate calculations on these sets of reactions will be presented and discussed. Parameters of modified Arrhenius type rate expressions fit to the calculated rate coefficients can be found in Supplementary Table S3.

### 3.1.1 Oxidation reactions involving H$_2$O

The reactions of SiO, Si$_2$O$_2$ and Si$_2$O$_3$ with H$_2$O are described schematically in the potential energy diagrams in Figures 3–5, respectively. Two things are apparent from these figures: (1) The reactions involve a multitude of pathways and reaction intermediates, and (2) there are high barriers to reaction to eventually lead to oxidized product plus $H_2$ (i.e., reactions 1a, 2a, and 3a). As noted in Section 2, the reactions were simplified by ignoring several high-energy pathways and by combining several intermediates that are close in structure and energy. Such combinations are indicated by the dashed lines in the potential energy diagram where several lower barriers and intermediates are left out. In the case of the $Si_2O_3 + H_2O$ reaction, there are six initial reaction paths (see Figure 5) that go from initial intermolecular complexes to the relatively stable intermediate species $H(Si_2O_3)OH$ (through TS1) or $Si_2O_2(OH)_2$ (through TS2—TS6). The former reaction is the only pathway that can eventually lead to the formation of $Si_2O_4 + H_2$ (reaction 4a) whereas the latter pathways either lead to formation of stabilized $Si_2O_2(OH)_2$ (reaction 4c) or formation of $SiO + SiO(OH)_2$ (reaction 4b). An analogous reaction of the latter reaction is also present for the $Si_2O_2 + H_2O$ reaction leading to the formation of $SiO + Si(OH)_2$ (reaction 3b). In effect these reactions lead to destruction of the larger $Si_2O_2$ and $Si_2O_3$ back into smaller species. The *destruction* reactions are endothermic whereas the oxidation reactions forming $H_2$ and $Si_2O_3$ and $Si_2O_4$ (reactions 3a and 4a), respectively, are somewhat exothermic. One would therefore expect the latter reactions to be more efficient. However, there are higher kinetic barriers leading to these exothermic reactions and as can be seen in Figure 6 the kinetics favors the destruction reactions that are several orders of magnitude faster than the oxidation reactions. Effectively, high abundances of $H_2O$ would therefore seem to counteract growth of larger silicon oxide species. The fate of the hydroxylated species, $Si(OH)_2$ and $SiO(OH)_2$, and whether they can initiate other reaction mechanisms is not taken into account here. Figure 7 shows the low-pressure rate constants for the recombination reactions forming the hydroxylated species $Si(OH)_2$ and $Si_2O_2(OH)_2$ (reactions 1b and 4c). The latter reaction has a large rate coefficient at low temperatures, but it decreases rapidly with increasing temperature. Goumans and Bromley (2012) identified this reaction (4c) as a key reaction in their suggested reaction scheme for the initial stages of silicate formation and even though our results do not directly contradict that assumption, the preceding reaction step would be the formation of $Si_2O_3$ from the reaction of $Si_2O_2$ and $H_2O$ (reaction 3a). Since this was shown to be inefficient above, some other mechanism for the formation of $Si_2O_3$ is needed for the reaction scheme to be valid.

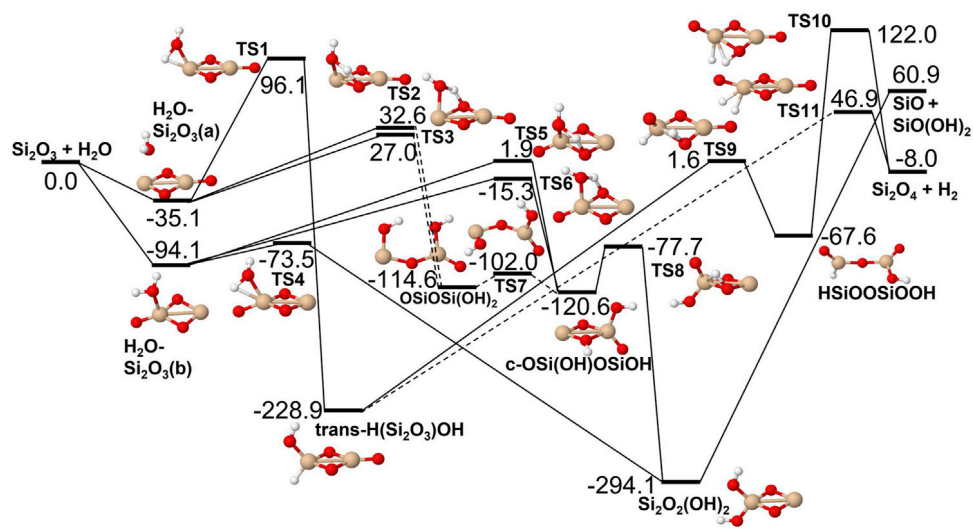

**FIGURE 5**
Potential energy surface based on M06 calculations for the $Si_2O_3 + H_2O$ reaction as implemented in the RRKM calculations. The energies are given in kJ/mol and include zero-point energy corrections. Dashed lines indicate that the reaction step has been simplified by combining several similar intermediate isomers.

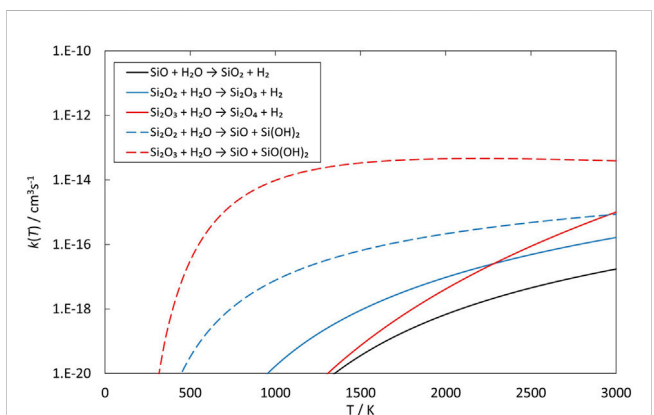

**FIGURE 6**
Rate coefficients (fit to RRKM calculations) of reactions involving SiO, $Si_2O_2$ and $Si_2O_3$ reacting with $H_2O$ as a function of temperature.

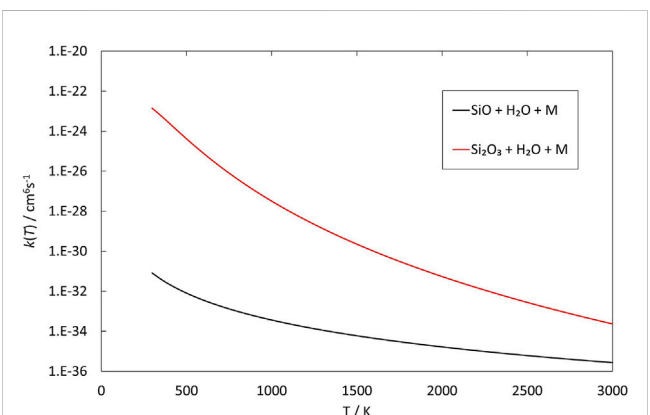

**FIGURE 7**
Rate coefficients (fit to RRKM calculations) of recombination reactions involving SiO and $Si_2O_3$ reacting with $H_2O$ as a function of temperature.

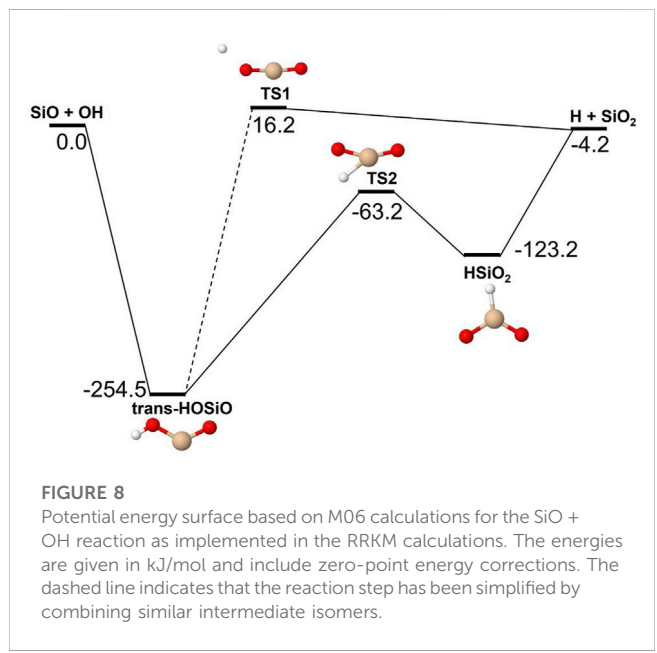

FIGURE 8
Potential energy surface based on M06 calculations for the SiO + OH reaction as implemented in the RRKM calculations. The energies are given in kJ/mol and include zero-point energy corrections. The dashed line indicates that the reaction step has been simplified by combining similar intermediate isomers.

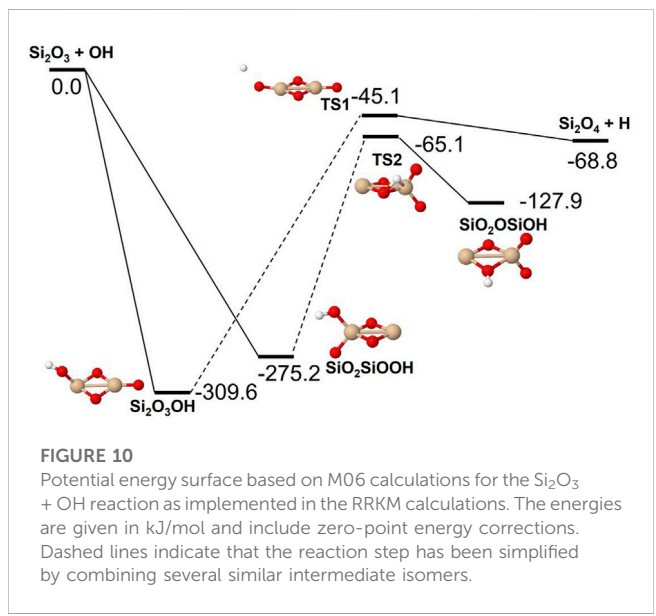

FIGURE 10
Potential energy surface based on M06 calculations for the $Si_2O_3$ + OH reaction as implemented in the RRKM calculations. The energies are given in kJ/mol and include zero-point energy corrections. Dashed lines indicate that the reaction step has been simplified by combining several similar intermediate isomers.

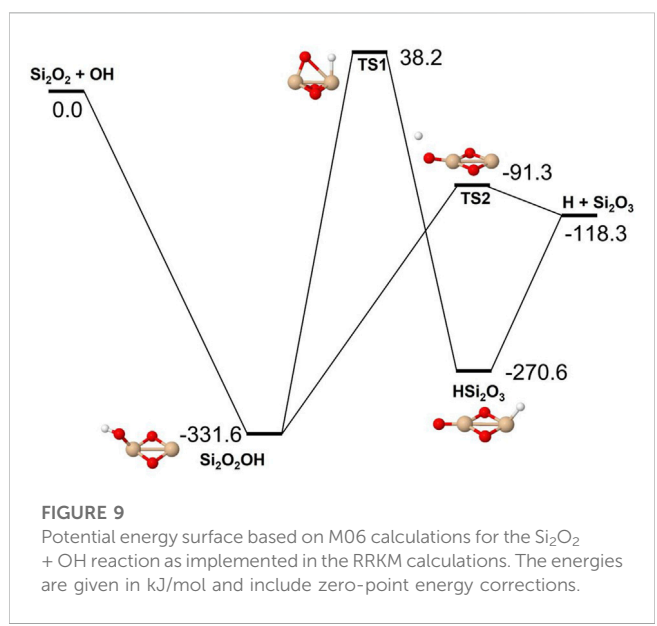

FIGURE 9
Potential energy surface based on M06 calculations for the $Si_2O_2$ + OH reaction as implemented in the RRKM calculations. The energies are given in kJ/mol and include zero-point energy corrections.

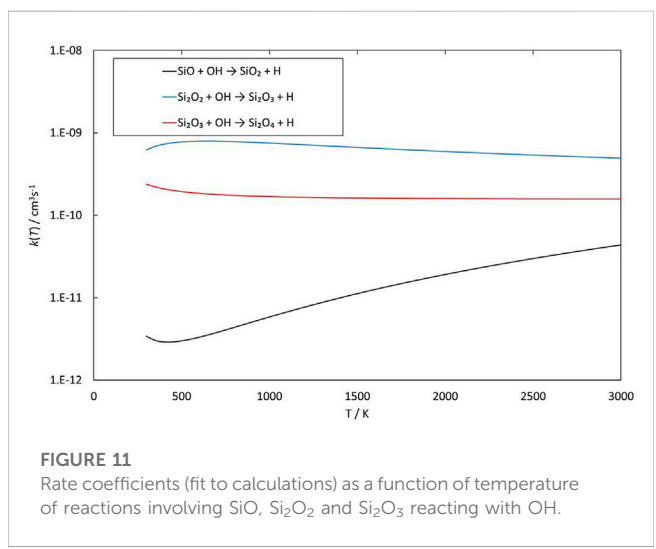

FIGURE 11
Rate coefficients (fit to calculations) as a function of temperature of reactions involving SiO, $Si_2O_2$ and $Si_2O_3$ reacting with OH.

### 3.1.2 Oxidation reactions involving OH

Reactions involving the reactive radical OH are expected to be significantly faster than the corresponding reactions with $H_2O$. Figures 8–10 show potential energy surfaces for the reactions of SiO, $Si_2O_2$, and $Si_2O_3$ with OH (reactions 5, 6, and 7) to form oxidized silicon oxides and H atoms, respectively. These reactions are all exothermic, but the exothermicity of the SiO + OH reaction is so small that the reaction is effectively thermoneutral. Comparing to the reactions with $H_2O$ it is also clear that the intermediate reaction barriers are much lower and that there are reaction paths that are lower in energy than the reactant energies. This suggests that the reactions should be fast. It should be noted that in the case of the $Si_2O_3$ + OH reaction described in Figure 10, the initial point of attack of OH is important for the final outcome. Only when OH reacts with

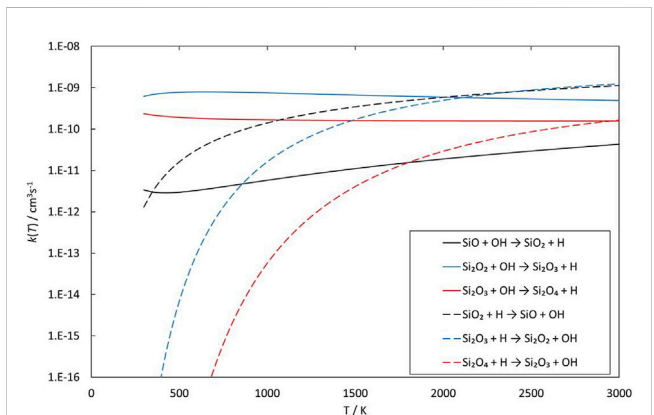

FIGURE 12
Rate coefficients (fit to calculations) as a function of temperature of reactions involving SiO, $Si_2O_2$ and $Si_2O_3$ reacting with OH and their reverse reactions.

TABLE 1 Reaction energies of clustering reactions calculated using M06/maug-cc-pV(T+d)Z calculations. The energies (at $T = 0$ K) are given in kJ/mol relative to the respective reactants and include zero-point energy corrections.

| Reactants | Product | M06 |
|---|---|---|
| SiO + SiO | $Si_2O_2$ | −231.3 |
| SiO + $SiO_2$ | $Si_2O_3$ | −345.4 |
| $SiO_2$ + $SiO_2$ | $Si_2O_4$ | −410.1 |
| SiO + $Si_2O_2$ | $Si_3O_3$ | −254.0 |
| SiO + $Si_2O_3$ | $Si_3O_4$ | −364.0 |
| $SiO_2$ + $Si_2O_2$ | $Si_3O_4$ | −478.1 |
| SiO + $Si_2O_4$ | $Si_3O_5$ | −383.3 |
| $SiO_2$ + $Si_2O_3$ | $Si_3O_5$ | −448.0 |
| $SiO_2$ + $Si_2O_4$ | $Si_3O_6$ | −453.0 |

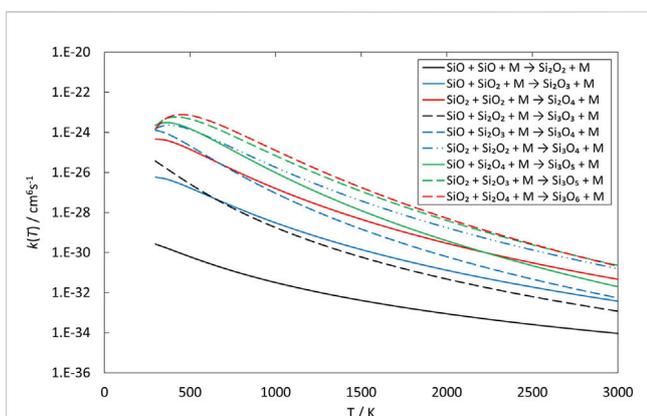

FIGURE 13
Rate coefficients (fit to RRKM calculations) of clustering reactions of small silicon oxide molecules as a function of temperature.

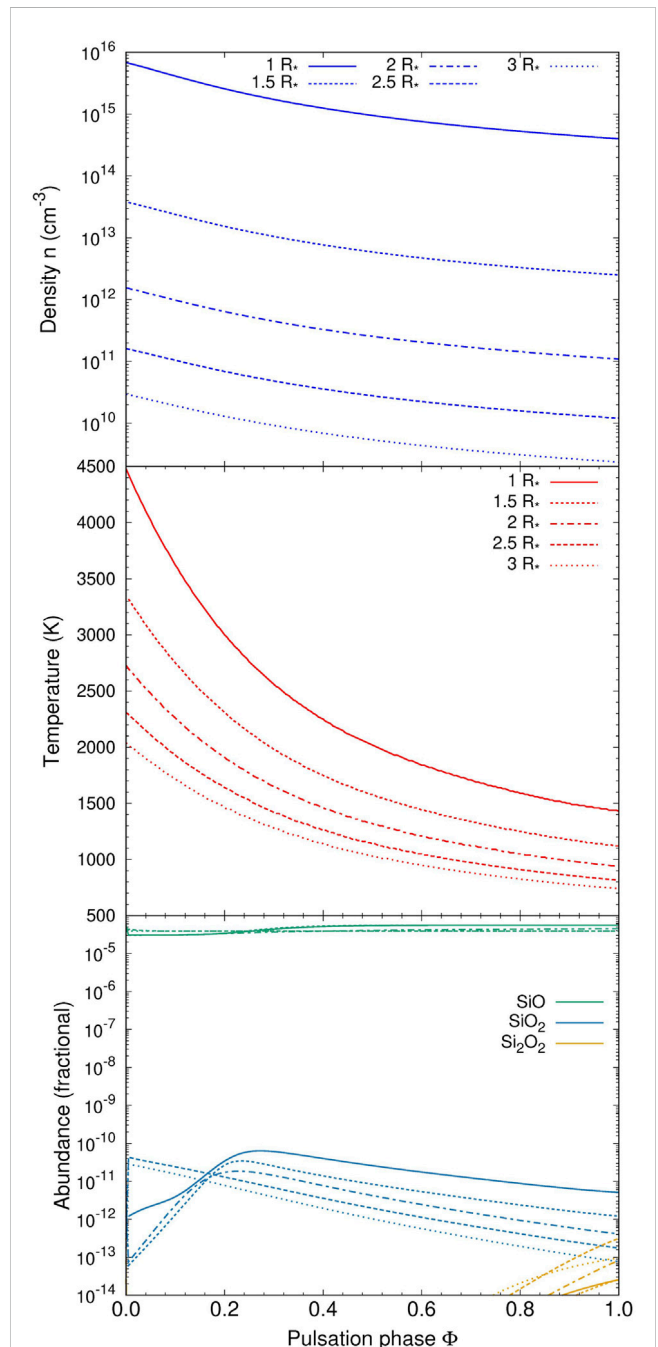

FIGURE 14
Density (top), temperature (middle), and fractional abundances in the kinetic model of an oxygen-rich AGB star as a function of pulsation phase (see text for details).

the Si atom that is bound to only two O atoms, the $Si_2O_4$ + H products can be formed. Reaction with the other Si atom, bound to three O atoms, does not lead to reaction or, more precisely, only leads to reaction through high-energy pathways with insignificant reaction rates (not shown in Figure 10). The calculated rate coefficients are shown in Figures 11, 12. In Figure 11, the forward reactions (involving OH) are presented. The reactions with OH are several orders of magnitude faster than the corresponding reactions with $H_2O$ (Figure 6), with the rate coefficients for $Si_2O_2$ and $Si_2O_3$ being basically temperature-independent and the one for SiO having a weakly positive temperature dependence. The reverse reaction of reaction 5, forming SiO + OH from H + $SiO_2$, is faster than the forward reaction over the whole temperature range considered, as also noted by Gómez Martín et al. (2009). The other reverse reactions,

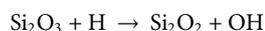

and

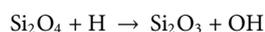

are slower than the forward reactions below about 2000 K and 3000 K, respectively, where the equilibrium is reversed, and they become faster instead. To efficiently form $Si_2O_3$ from $Si_2O_2$, as required in the scheme by Goumans and Bromley (2012), would be possible if a significant fraction of $H_2O$ has been dissociated into OH, but temperature should then preferably not be too high.

### 3.1.3 Clustering reactions

The clustering reactions (reactions 8—16) are all treated as a simple direct recombination of species in a single reaction step, i.e., without intermediates or reaction barriers. Some studies have suggested that there are intermediates and/or small barriers to formation of for instance $Si_3O_3$ (reaction 11, Avramov et al., 2005; Pimentel et al., 2006), but we follow Bromley et al. (2016) who concluded that any reaction barriers would be below the reactants and would not appreciably affect the kinetics of SiO clustering reactions. The energetics of the reactions are summarized in Table 1 and calculated rate coefficients (in the low-pressure limit) are shown in Figure 13. All rate coefficients decrease with temperature above 500 K but a few of the reactions are most efficient around 400—500 K. Most of the rate coefficients are quite similar in magnitude and it is only reaction 8, the formation of $Si_2O_2$ from two SiO molecules, that is clearly slower than the other reactions. As mentioned earlier, this is one of the key reactions in the silicate formation scheme by Goumans and Bromley (2012), and it does clearly seem to be a bottleneck for further growth of silicon oxides.

## 3.2 Kinetic modeling in an oxygen-rich circumstellar environment

For the hydrodynamic trajectories pertaining to the circumstellar environment of an oxygen-rich AGB star we use the pulsating MIRA model reported in Gobrecht et al. (2022). This model is characterized by a C/O ratio of 0.75, an effective temperature of $T = 2000$ K and a photospheric density at 1R* of $n = 4 \times 10^{14}$ cm$^{-3}$. The computational domain extends from 1 to 3 stellar radii and is subject to pulsation-induced shocks that periodically cross the atmosphere with a period of $p = 470$ days. The resulting post-shock density and temperature profiles are shown in the top and middle panel of Figure 14 as a function of pulsation phase and radial distance.

The chemistry is modeled by a kinetic rate network that includes the prevalent gas phase species $H_2$, CO, $CO_2$, $H_2O$, OH, and SiO, whose initial pre-shock abundances at R = 1R* are given by thermodynamic equilibrium. The corresponding reaction rates are listed as reactions 1–25 in Gobrecht et al. (2022). In addition, rates linking the $Si_xO_yH_z$, x = 1—3, y = 2—6, z = 0—2, clusters are an integral part of the present study, and these were included in our chemical-kinetic network (see Supplementary Table S3). Kinetic parameters were taken from fits to the RRKM calculations described in Section 3.1, except for those for reaction 2 (Plane et al., 2016). The integration of differential rate equations was performed using the Linear Solving of Ordinary Differential Equations (LSODE) solver (Hindmarsh, 1980).

In the bottom panel of Figure 14, the fractional abundances of silicon-, and oxygen-bearing molecules and clusters are shown. The chemistry is dominated by the SiO molecule for all considered radial distances and pulsation phases. $SiO_2$ is formed in modest quantities peaking at phases 0.25—0.3 for radial distances ≤2 R*. In these early post-shock regions, some OH is available and small concentrations of $SiO_2$ are synthesized by its predominant formation channel SiO + OH. The $Si_2O_2$ cluster forms only in tiny amounts at late phases (>0.8). All other considered $Si_xO_yH_z$ species show negligible abundances with values below $10^{-14}$ in the entire computational domain.

We did not include alternative reaction mechanism that could be important, most prominently radiative association. Plane and Robertson (2022) showed that for the $SiO_2$ + $H_2O$ recombination reaction, radiative association becomes the dominant mechanism for densities lower than $10^9$ cm$^{-3}$, which is five orders of magnitude lower than the unshocked photospheric density used in our study. At this and higher densities termolecular recombinations dominate over radiative association.

## 4 Conclusion

DFT and RRKM calculations were performed for a number of reactions involving SiO, $Si_2O_2$, and $Si_2O_3$. Rate coefficients were evaluated in the temperature range 298–3000 K. Reactions involving $H_2O$ do not efficiently form the oxidized forms of the silicon oxides, i.e., $SiO_2$, $Si_2O_3$, and $Si_2O_4$, respectively, but might lead to formation of hydroxylated species instead. The reaction of $Si_2O_3$ with $H_2O$ to form $Si_2O_2(OH)_2$ seems to be particularly efficient under the right conditions, i.e., low temperatures and sufficiently high pressure. In contrast, reactions with OH radicals are fast at all temperatures and lead to formation of $SiO_2$, $Si_2O_3$, and $Si_2O_4$. Especially at high temperatures the reverse reactions become faster than the forward reactions, potentially decreasing the efficiency of formation of the oxidized silicon oxide species. The clustering reactions leading to build-up of larger silicon oxide species are all most efficient at low temperature. A reaction scheme for the initial stages of circumstellar silicate formation proposed by Goumans and Bromley (2012) based on reaction thermodynamics, and starting from SiO, $H_2O$ and Mg, depends on the efficient formation of $Si_2O_2$ from clustering of two SiO molecules, formation of $Si_2O_3$ from the reaction of $Si_2O_2$ with $H_2O$, and subsequently formation of $Si_2O_2(OH)_2$ from the reaction of $Si_2O_3$ and $H_2O$. Out of these three reaction steps, the efficiency of $Si_2O_3$ formation is not supported by our results. However, if there would be sufficiently high abundances of OH, for instance through thermal or photo-induced dissociation of $H_2O$, the reaction might still be sufficiently efficient for the reaction scheme to be valid. Kinetic modeling of the outflows from AGB stars based on our results lead to only minor amounts of $SiO_2$ and $Si_2O_2$ being formed.

This seems to indicate that silicon oxide molecules alone will not initiate growth of larger oxide particles in stellar outflows.

## Data availability statement

The original contributions presented in the study are included in the article/Supplementary Material, further inquiries can be directed to the corresponding author.

## Author contributions

SA performed the DFT and RRKM calculations and wrote most of the manuscript. DG performed the kinetic modeling of AGB stars, wrote the corresponding part and edited parts of the introduction. RV provided valuable input to the study of the SiO + OH reaction.

## Funding

The Funding from Knut and Alice Wallenberg foundation (research grant KAW 2020.0081) is gratefully acknowledged.

## Conflict of interest

RV was employed by Zhejiang Huayou Cobalt Co., Ltd.

The remaining author declares that the research was conducted in the absence of any commercial or financial relationships that could be construed as a potential conflict of interest.

## Publisher's note

All claims expressed in this article are solely those of the authors and do not necessarily represent those of their affiliated organizations, or those of the publisher, the editors and the reviewers. Any product that may be evaluated in this article, or claim that may be made by its manufacturer, is not guaranteed or endorsed by the publisher.

## Supplementary material

The Supplementary Material for this article can be found online at: https://www.frontiersin.org/articles/10.3389/fspas.2023.1135156/full#supplementary-material